\documentclass[aps,prd,nofootinbib,floatfix,twocolumn]{revtex4}
\usepackage{amsmath,amssymb,color,graphicx,bm}
\usepackage{physics}

\definecolor{red}{rgb}{0.8,0,0}
\definecolor{RED}{rgb}{0.8,0,0}
\definecolor{violet}{rgb}{0.4,0,0.4}
\definecolor{green}{rgb}{0,0.5,0.0}
\definecolor{GREEN}{rgb}{0,0.5,0.0}
\definecolor{navy}{rgb}{0.0,0.0,0.6}
\definecolor{orange}{rgb}{0.8,0.2,0.0}
\definecolor{blue}{rgb}{0.3,0.0,0.8}

\begin{document}

\title{\textbf{Quantum ballet by gravitational waves: Generating entanglement's dance of revival-collapse and memory within the quantum system}}
\author{Partha Nandi$^a$}
\email{pnandi@sun.ac.za}
\author{Bibhas Ranjan Majhi$^b$}
\email{bibhas.majhi@iitg.ac.in}
\author{Nandita Debnath$^c$}
\email{debnathnandita14@gmail.com}
\author{Subhajit Kala$^b$}
\email{s.kala@iitg.ac.in}
\affiliation{$^a$Department of Physics, University of Stellenbosch, Stellenbosch-7600, South Africa\\
$^b$Department of Physics, Indian Institute of Technology Guwahati, Guwahati 781039, Assam, India.\\
$^c$Department of Physics, University of Calcutta, Kolkata 700009, India.}


\begin{abstract}

Recent proposals are emerging for the experimental detection of entanglement mediated by classical gravity, carrying significant theoretical and observational implications. In fact, the detection of gravitational waves (GWs) in LIGO provides an alternative laboratory for testing various gravity-related properties. By employing LIGO's arms as oscillators interacting with gravitational waves (GWs), our study demonstrates the potential for generating quantum entanglement between two mutually orthogonal modes of simple harmonic oscillators. Our findings reveal unique entanglement dynamics, including periodic ``collapse and revival" influenced by GW oscillations, alongside a distinct ``quantum memory effect." Effectively, each harmonic oscillator feels a temperature. We believe that these forecasts may hold significance towards both theoretically probing and experimentally verifying various properties of classical gravitational waves.

\end{abstract}
\maketitle	

\section{Introduction}\label{Sec1}
Ever since the development of quantum mechanics and
general relativity in the first decades of the last century,
the construction of a successful quantum theory of gravity
poses a problem to this day. Moreover, we do not have
experimental results to guide such a construction since
quantum gravitational effects are expected to be relevant
at the Planck scale, which is far beyond our current technology.
Recently, there has been growing interest for the laboratory detection of entanglement mediated by gravity \cite{Bose:2017nin,Marletto:2017kzi,Danielson:2021egj}. The theoretical proposal has been based on an information–theoretic argument --  the classical field can not establish entanglement between the quantum states of two spatially separated objects. Consequently, there have been proposals for tabletop experiments to detect such entanglement \cite{Bose:2017nin,Marletto:2017kzi,Christodoulou:2022mkf}.

One of the predictions from Einstein’s general theory of
relativity is the phenomenon of gravitational waves (GWs) \cite{1,2}.
Experiments to detect gravitational waves began with Weber and his resonant mass detectors in the 1960s \cite{PhysRev.117.306}. 
In recent years, the ground-breaking detection of GWs by the LIGO detector in 2015 \cite{LIGOScientific:2016aoc,LIGOScientific:2017vwq,LIGOScientific:2020iuh}, which was
hitherto inaccessible to us, has opened up as a new data
source for astronomical as well as cosmological inference \cite{PhysRevD.101.122001,Willke:2006uw,Wang:2019tto}. This not only provides us with an observational
probe into some strong-gravity scenarios, like black-hole
and neutron star mergers, but also with their enhanced
signal-to-noise ratio the upcoming ground-based GWs
observatories like the Einstein telescope \cite{amaroseoane2017laser}, the Cosmic Explore \cite{Punturo:2010zz} and the space-based LISA mission \cite{LIGOScientific:2016wof}
can uncover signatures of new physics.

There has been a surge of works in investigating how classical gravitational waves influence quantum phenomena \cite{PhysRevD.51.1701,PhysRevD.107.063503}. Instead of focusing on the quantum nature of gravity itself, experiments like the COW experiment \cite{PhysRevLett.34.1472} and investigations conducted by Pikovski et al. \cite{Pikovski:2013qwa} delve into studying the quantum aspects of particles under classical external gravity. This revived curiosity has led to exploring gravitationally mediated entanglement \cite{Bose:2017nin,Marletto:2017kzi,Danielson:2021egj}, the implications of spacetime superposition \cite{Christodoulou:2022knr}, and examining causal order within gravitational scenarios \cite{Christodoulou:2022mkf}. Additionally, there is an increasing curiosity surrounding the investigation of gravitational decoherence induced by weak gravitational fields, observed in both classical and quantum frameworks. A detailed overview of this area, complete with pertinent references, is available in Ref \cite{Blencowe:2012mp,Pikovski:2013qwa}. Within the realm of gravitational decoherence, gravitational perturbations are perceived as an environment that affects the progression of quantum particles existing in superposition within a defined basis, ultimately resulting in the dissipation of quantum coherence. Recently, we observed that low-frequency linearized GWs are capable of inducing the Berry phase in a state of a quantum system \cite{Nandi:2022sjy} (readdressed in \cite{Sen:2023vor}). Although it may not offer a complete description of quantum gravitational effects, it could shed light on the necessity (or lack thereof) to quantize the field \cite{PhysRevD.108.L101702}. Simultaneously, it provides valuable experimental insights for the theoretical construction of the overarching theory. Therefore in the search for a laboratory experiment, the GWs may open up an alternative laboratory, particularly to witness a detection of aforesaid entanglement phenomenon, just much like black hole can be laboratory to venture quantum aspects of strong gravity.


Here we provide a very simple quantum mechanical model to theoretically witness such entanglement. This model is very much based on the LIGO apparatus. Therefore provides a possibility to perform the experiment with an ``upgraded version'' of LIGO in the near future. Since the length of the two perpendicular arms of the LIGO is very small compared to the wavelength of the GWs, the endpoints of each arm can be treated as a point particle that can perform two-dimensional oscillations. We consider each of them to be a two-dimensional harmonic oscillator (HO) in the absence of GWs. Now we know that in the presence of the GWs (at the linearized theory), the geodesic of a particle suffers deviations along the directions perpendicular to the propagation of the GWs. Therefore the dynamics of these two perpendicular oscillators will be affected when interacting with GWs. Treating the GWs classically and the oscillators as quantum HOs, we find an effective Hamiltonian of the system, the full classical counterpart of which already exists in the literature. In a particular choice of canonical variables for the particles, we observe that two HOs, which were independent in the absence of GWs, are now coupled with each other through the contribution from GWs. The identified interaction term in terms of creation and annihilation operators of the uncoupled HOs represents a two-mode squeezing operator. 

Treating this as a small perturbation on the uncoupled HOs system, we compute the time-evolved quantum state of the system's initial state. Particularly, we consider the initial composite state as the ground state of each HO. We observe that the later time state can no longer be an eigenstate of individual number operators of the HOs. Furthermore, depending on the values of the parameter space\textbf{,} the initial pure state of each HO evolves to a mixed state thereby producing entanglement \cite{RevModPhys.81.865} between the quantum states of the two HOs. Our focus in this study lies in exploring the dynamic behaviour of this GWs mediated entanglement between independent modes within a two-dimensional isotropic oscillator detector affected by gravitational waves. To quantify the entanglement and mixing, we calculate von Neumann entropy and Purity till second order in perturbation expansion and find that these depend on the nature of GWs and the system's frequencies.  

We find two novel features in this investigation. For a sinusoidal interaction produced by periodic GWs, the quantity (we call as $|c_+(t)|^2$) that quantifies the dynamical behaviour of entanglement shows two very distinct properties. We observe that there are instances at which the interaction is switched off, but the entanglement is non-vanishing. Therefore, once they are entangled by the GWs, they remain quantum mechanically connected even if they do not suffer classical interaction. This we call as the ``{\it quantum memory effect}''. Another interesting characteristic is that for a few consecutive maxima of the interaction, the entanglement increases and reaches to a peak. After that, for the next few maxima, it decreases. The same is repeated periodically. This dynamical behaviour is called as the entanglement ``{\it collapsing and reviving}'' phenomenon \cite{PhysRevLett.65.3385,PhysRevLett.65.3385+}.
Notably, the fluctuating nature of entanglement, characterized by the periodic fluctuation of entanglement, has garnered significant attention in recent years within bipartite systems involving quantum optics and laser physics. Moreover, we find that each HO registers the interaction with the GWs as it is immersed in an effective thermal bath.




\section{The model: GWs squeezes quantum system}\label{Sec2}
The deviation of a geodesic in the presence of linearized GWs can be thought of as a motion of a particle in two dimensions, perpendicular to the propagation of GWs \cite{Maggiore:2007ulw,Misner:1973prb}. In LIGO, the endpoint of each arm, which can be regarded as a point particle, will then execute two-dimensional motion under the influence of GWs. In such a picture, if the detector is kept in an external harmonic potential, then the Hamiltonian for each arm can be taken as \cite{Nandi:2022sjy,PhysRevD.104.046021}
\begin{equation}
H = \sum_{j=1,2} \Big(\frac{P_j^2}{2m_{0}} + \sum_{k=1,2} \Gamma^j_{0k} X^k P_j + \frac{1}{2} m_{0}\Omega_{0}^2X_j^2\Big)~.
\label{H1}
\end{equation}
Here, $m_{0}$ denotes the test particle's mass, $\Gamma^j_{0k}$ represents the Christofell connection coefficients, and $\Omega_{0}$ is the frequency corresponding to the harmonic potential. The model can be envisioned as two harmonic oscillators perturbed by gravitational waves (GWs), a concept previously explored \cite{PhysRevD.51.1701,PhysRevD.97.044015}.

For the chosen GWs $h_{jk}(t) = 2\chi(t)(\epsilon_{\times}\sigma_{1jk} + \epsilon_{+}\sigma_{3jk})$,  the quantum mechanical equivalent of the aforementioned Hamiltonian takes on the following expression \cite{Nandi:2022sjy}:
\begin{eqnarray}
\hat{H}(t)&=&\sum_{i=1,2}\Big(\alpha \hat{P}_{i}^2+\beta \hat{X}_{i}^2\Big)+\gamma(t)(\hat{X}_{1}\hat{P}_{1}+\hat{P}_{1}\hat{X}_{1})
\nonumber
\\
&&-\gamma(t)(\hat{X}_{2}\hat{P}_{2}+\hat{P}_{2}\hat{X}_{2})+\delta(t)(\hat{X}_{1}\hat{P}_{2}+\hat{P}_{1}\hat{X}_{2})~,
\label{BP1}
\end{eqnarray}
where $\alpha = \frac{1}{2m_{0}}$, $\beta = \frac{1}{2} m_{0}\Omega_0^2$, $\gamma(t) = \frac{\dot{\chi}(t)\epsilon_+}{2}$ and $\delta(t) = \dot{\chi}(t)\epsilon_{\times}$. Here $2\chi(t)$ denotes the time variation of the GW. $\sigma_{1jk}$ is the $(jk)^{th}$ element of the Pauli matrix $\sigma_1$, and so on.

For our future purpose, in order to simplify the calculation, we will work on a new set of canonical variables. These are achieved by applying a phase space canonical transformations of the form:
\begin{equation}
\begin{pmatrix}
\hat{X}_{1}\\
\hat{X}_{2}\\
\end{pmatrix}\rightarrow\begin{pmatrix}
\hat{X}_{1}^\prime \\
\hat{X}_{2}^\prime\\
\end{pmatrix}=\begin{pmatrix}
\cos{\theta} & \sin{\theta}\\
-\sin{\theta} & \cos{\theta}\\
\end{pmatrix}\begin{pmatrix}
\hat{X}_{1}\\
\hat{X}_{2}\\
\end{pmatrix}~;
\end{equation}
and
\begin{equation}
\begin{pmatrix}
\hat{P}_{1}\\
\hat{P}_{2}\\
\end{pmatrix}\rightarrow\begin{pmatrix}
\hat{P}_{1}^\prime \\
\hat{P}_{2}^\prime\\
\end{pmatrix}=\begin{pmatrix}
\cos{\theta} & \sin{\theta}\\
-\sin{\theta} & \cos{\theta}\\
\end{pmatrix}\begin{pmatrix}
\hat{P}_{1} \\
\hat{P}_{2}\\
\end{pmatrix}~.
\end{equation}
Then with a specific choice of $\theta$, given by $\tan{2\theta}=-\frac{2\gamma(t)}{\delta(t)}$, one finds the desired form of the Hamiltonian as
\begin{equation}
\hat{H}(t)=\sum_{i=1,2}\Big(\alpha {{\hat{P}}_{i}^{\prime^2}}+\beta {\hat{X}_{i}^{\prime^2}}\Big) + g(t)({\hat{X}_{1}^{\prime}}{\hat{P}_{2}^{\prime}}+{\hat{P}_{1}^{\prime}}{\hat{X}_{2}^{\prime}})~, 
\label{BRM1}
\end{equation}
where $g(t) = \pm {\sqrt{\delta^2(t)+4\gamma^2(t)}}$.

Since the above Hamiltonian is already in Hermitian form, the introduction of the annihilation and creation operators for the two-mode harmonic oscillator, defined by
\begin{eqnarray}
&&\hat{a}_{i}=\Big(\frac{\alpha}{\beta}\Big)^{1/4}\Big(\frac{\sqrt{\frac{\beta}{\alpha}}\hat{X}_{i}^{\prime}+i\hat{P}_{i}^{\prime}}{\sqrt{2\hbar}}\Big)~;
\nonumber
\\
&&\hat{{a}}_{i}^{\dagger}=\Big(\frac{\alpha}{\beta}\Big)^{1/4}\Big(\frac{\sqrt{\frac{\beta}{\alpha}}\hat{X}_{i}^{\prime}-i\hat{P}_{i}^{\prime}}{\sqrt{2\hbar}}\Big)~,
\end{eqnarray}
with $[\hat{a}_{i},\hat{{a}}_{j}^{\dagger}]=\delta_{ij}$, leads to  
\begin{equation}
\hat{H}(t)=2\hbar\sqrt{\alpha\beta}\Big(\sum_i \hat{N}_i+1\Big)+i\hbar g(t)(\hat{{a}}_{1}^{\dagger}\hat{{a}}_{2}^{\dagger}-\hat{a}_{1}\hat{a}_{2})~.
\label{BRM3}
\end{equation}
Here we denote $\hat{N}_i = \hat{{a}}_{i}^{\dagger}\hat{a}_{i}$.
Note that the two-dimensional HO is now affected by the GWs and the modification in the Hamiltonian is encoded in the term containing $g(t)$. In the absence of GWs, $g(t)$ vanishes and its presence provides a two-mode squeezing term. Considering this term as a perturbation, we expect that such a squeezing will have a very prominent effect on the quantum states of the HO.

In order to engage in the quantum calculation, we will first study the group structure associated with the Hamiltonian. For that purpose we define following operators: 
\begin{equation}
\hat{K}_{0}=\frac{1}{2}\Big(\sum_i\hat{N}_i+1\Big);~~\hat{K}_{+}=\hat{{a}}_{1}^{\dagger}\hat{{a}}_{2}^{\dagger};~~\hat{K}_{-}=\hat{a}_{1}\hat{a}_{2}~,
\label{BRM2}
\end{equation}
which satisfies the algebra corresponding to a two-mode or spinor representation of $SU(1,1)$ group structure, given by
\begin{equation}
[\hat{K}_{0},\hat{K}_{\pm}]=\pm\hat{K}_{\pm};~~~~[\hat{K}_{+},\hat{K}_{-}]=-2\hat{K}_{0}~.
\label {su}
\end{equation}
Hence the operators, defined in (\ref{BRM2}), are the generators of  $SU(1,1)$ group.
In terms of these generators, (\ref{BRM3}) reduces to the following form:
\begin{equation}
\hat{H}(t)=\hbar\Omega\hat{K}_{0}+i\sigma(t)(\hat{K}_{+}-\hat{K}_{-})~,
\label{BRM4}
\end{equation}
with $\Omega=4\sqrt{\alpha\beta}=2\Omega_0$ and $\sigma(t)=\hbar g(t)$. Note that $K_0^\dagger=K_0,~K_{+}^\dagger=K_{-}$ and $K_{-}^\dagger=K_{+}$, which we will use in various stages.


\section{Entanglement phenomenon}\label{Sec3}
We consider the interaction between the modes of the HO and the GWs as very weak so that the second term in (\ref{BRM4}) can be treated as a perturbation. In this case (\ref{BRM4}) can be treated as $\hat{H} = \hat{H}_0 + \hat{H}_{\textrm{int}}$, where $\hat{H}_{\textrm{int}} = i\sigma(t)(\hat{K}_{+}-\hat{K}_{-})$. Since it is time-dependent, the system can be studied with time-dependent perturbation theory, in which the interaction picture is the suitable formalism. The whole analysis will be done within the second-order perturbation calculation, i.e., up to order $g^2(t)$ and therefore the terms with ($\sim\sigma^3$) and onwards will be neglected. We will find that retaining up to quadratic terms in the perturbation series leads to a leading-order non-trivial contribution in our desired quantities.

Consider that the two-dimensional HO system initially ($t=0$) was in the ground state $\ket{0,0; t=0}$.
Then in the interaction picture, the time evolution of the state is given by $\ket{0,0;t}_I=\hat{U}_I(t,0)\ket{0,0;t=0}$, where 
\begin{equation}
{\hat U}_I(t,0)={\hat T}e^{-\frac{i}{\hbar}{\int_{0}^{t}{\hat {H}}_{int}^I(t^\prime)dt^\prime}}~. 
\end{equation}
Here ${\hat T}$ represents the time-ordered product between the operators. In the above, ${\hat {H}}_{int}^I$ is given by 
\begin{eqnarray}
\hat{H}_{int}^I(t)&=&e^{\frac{i\hat{H}_0t}{\hbar}}\hat{H}_{int}(t)e^{-\frac{i\hat{H}_0t}{\hbar}}
\nonumber
\\
&=&i\sigma(t) e^{i\Omega \hat{K}_{0}t}(\hat{K}_{+}-\hat{K}_{-})e^{-i\Omega \hat{K}_{0}t}~.
\end{eqnarray}
Using the well-known Baker-Campbell-Hausdorff (BCH) lemma, we get a simplified form of $\hat{H}_{int}^I(t)$ as
\begin{equation}
\hat{H}_{int}^I(t)=i\sigma(t)(e^{i\Omega t}\hat{K}_{+}-e^{-i\Omega t}\hat{K}_{-})~.
\end{equation}
Then employing the Magnus expansion formula, ${\hat U_I(t,0)}$ can be computed as \cite{Magnus:1954zz}
\begin{eqnarray}
&&\hat U_I(t,0)=\exp\Big(-\frac{i}{\hbar}{\int_{0}^{t}{\hat {H}}_{int}^I(t_1)dt_1}
\nonumber
\\
&&-\frac{1}{2!\hbar^2}{\int_{0}^{t}dt_1{\int_{0}^{t_1}dt_2[{\hat {H}}_{int}^I(t_1),{\hat {H}}_{int}^I(t_2)]}} + \mathcal{O}(\sigma)^3\Big)~.
\label{teo}
\end{eqnarray}
Now considering the fact $[{\hat {H}}_{int}^I(t_1),{\hat {H}}_{int}^I(t_2)]=-4i\sigma(t_1)\sigma(t_2)\hat{K}_0\sin{[\Omega(t_1-t_2)]}$ and keeping upto $\sigma^2$ order, one finds
\begin{equation}
\hat U_I(t,0)= 1 + i(c_0\hat{K}_0 - c_+\hat{K}_+ +  c_-\hat{K}_-) - \frac{1}{2}(c_+\hat{K}_+ -  c_-\hat{K}_-)^2~,
\label{BRM6}
\end{equation}
where $c_+(t)=\int_{0}^{t}\frac{i\sigma(t_1)}{\hbar}e^{i\Omega t_1}dt_1$ and $c_-(t)=-\bar{c}_{+}(t)$, $c_{0}(t)=\frac{2}{\hbar^2}\int_{0}^{t}dt_1\int_{0}^{t_1}dt_2\sigma(t_1)\sigma(t_2)\sin{[\Omega(t_1-t_2)]}$. 

Now we proceed to find the final state. Inserting the completeness relation for eigenstates of $\hat{H}_0$\textbf{,} we write
\begin{equation}
\ket{0,0;t}_I=\sum_{m_1,m_2=0}^\infty d_{m_1,m_2}^{(0,0)}(t) \ket{m_1,m_2;t=0}~,
\label{wig}
\end{equation}
where
\begin{equation}
    d_{m_1,m_2}^{(0,0)}=\bra{m_1,m_2;t=0}\hat{U}_I(t,0)\ket{0,0;t=0}~.
\end{equation}
Using the facts $\hat{N}_1\ket{0,0;t=0}=0=\hat{N}_2\ket{0,0;t=0}$ and $[\hat{N}_1-\hat{N}_2, \hat{U}_I(t,0)]=0$, one finds $(\hat{N}_1-\hat{N}_2)\ket{0,0;t}_I = (\hat{N}_1-\hat{N}_2)\hat{U}_I(t,0)\ket{0,0;t=0}=0$. Application of this in (\ref{wig}) yields
\begin{equation}
\sum_{m_1,m_2=0}^\infty d_{m_1,m_2}^{(0,0)}(t)(m_1-m_2)\ket{m_1,m_2;t=0}=0~.
\end{equation}
Then one finds $d_{m_1,m_2}^{(0,0)}(t)=d_{m_1}(t)\delta_{m_1m_2}$ and so (\ref{wig}) reduces to
\begin{equation}
\ket{0,0;t}_I=\sum_{m=0}^\infty d_{m}(t) \ket{m,m;t=0}~.
\label{BRM5}
\end{equation}
As $[\hat{N}_1-\hat{N}_2, \hat{U}_I(t,0)]=0$ and $\hat{U}_I(t,0)$ is unitary, one can define two operators $\hat{{b}}_{i}=\hat U_I\hat{{a}}_{i}\hat{U}_I^\dagger$ such that $\hat{N}_1-\hat{N}_2 = \hat{{b}}_{1}^{\dagger}\hat{b}_{1}-\hat{{b}}_{2}^{\dagger}\hat{b}_{2}$ and $\hat{b}_i \ket{0,0;t}_I = 0$. For our unitary operator (\ref{BRM6}), we find
\begin{equation}
\hat{b}_1(t) = A_1\hat{a}_1+A_2\hat{a}_2^{\dagger}~; \,\,\
\hat{b}_2(t) = A_1\hat{a}_2+A_2\hat{a}_1^{\dagger}~,
\label{BRM7}
\end{equation}
where $A_1(t) = 1+\frac{|c_+|^2}{2} - \frac{ic_0}{2}$ and $A_2(t) = ic_+$. Using the above results in $\hat{b}_i \ket{0,0;t}_I = 0$ along with (\ref{BRM5}), one obtains a recursion relation $d_m(t) = -\frac{A_2}{A_1}d_{m-1}(t)$ for $m\geq 1$. This yields $d_m(t)=\Big(-\frac{A_2}{A_1}\Big)^md_{0}(t)$ with $m\geq 1$. $d_0$ can be determined by demanding the normalization of $\ket{0,0;t}_I$, given by (\ref{BRM5}); i.e. from $\sum_{m=0}^\infty |d_m|^2 = 1$. Using this, one finds $d_m = \Big(-\frac{A_2}{A_1}\Big)^m \sqrt{1-\Big|\frac{A_2}{A_1}\Big|^2}$ for $m\geq 1$. Then the normalized final state is given by
\begin{equation}
\ket{0,0;t}_I = \sqrt{1-\Big|\frac{A_2}{A_1}\Big|^2} \sum_{m=0}^\infty \Big(-\frac{A_2}{A_1}\Big)^m \ket{m,m;t=0}~.
\label{BRM8}
\end{equation}
The reduced density matrix of one of the oscillators (say, the first oscillator) is then obtained as
\begin{eqnarray}
&&\hat{\rho}_1(t) = \textrm{Tr}_2 (\hat{\rho}(t)) = \textrm{Tr}_2 \Big({\ket{0,0;t}_I} {\bra{0,0;t}}\Big)
\nonumber
\\
&&= \Big(1-\Big|\frac{A_2}{A_1}\Big|^2\Big) \sum_{m=0}^\infty \Big|\frac{A_2}{A_1}\Big|^{2m} \ket{m;t=0}\bra{m;t=0}~.
\label{BRM9}
\end{eqnarray}
In the above, $\textrm{Tr}_2$ represents the tracing over the quantum states of the second oscillator.
In order to investigate whether the final state (\ref{BRM8}) represents an entangled state, we will be calculating two quantities: von Neumann entropy and purity, which will quantify the generation of entanglement between the two HOs.

The later time entropy (von-Neumann) of the first oscillator can be determined by the formula $S(t) = -\textrm{Tr}_1\Big(\hat{\rho_1}(t)\ln\hat{\rho}_1(t)\Big)$. From (\ref{BRM9})\textbf{,} this is evaluated as
\begin{eqnarray}
S(t) = -\ln\Big(1-\Big|\frac{A_2}{A_1}\Big|^2\Big) - \frac{\Big|\frac{A_2}{A_1}\Big|^2\ln\Big|\frac{A_2}{A_1}\Big|^2}{\Big(1-\Big|\frac{A_2}{A_1}\Big|^2\Big)}~.
\label{BRM10}
\end{eqnarray}
Using the explicit values of $A_1$ and $A_2$, one finds till $\mathcal{O}(\sigma^2)$ as $\Big|\frac{A_2}{A_1}\Big|^2 \simeq |c_+|^2$. Then the entropy turns out to be
\begin{equation}
S(t) \simeq \Big|c_+(t)\Big|^2~.
\label{BRM11}
\end{equation} 
Note that \textbf{$S(t)$} can be non-vanishing, although it was zero at $t=0$. This indicates that the GWs assist a quantum entanglement between our two HOs, whose initial states were not entangled.

The Purity function is defined as $P(t) = \textrm{Tr}_1(\hat{\rho}_1^2)$. This in our case is calculated as
\begin{equation}
P(t) = \frac{1-\Big|\frac{A_2}{A_1}\Big|^2}{1+\Big|\frac{A_2}{A_1}\Big|^2} \simeq 1 - 2\Big|c_+(t)\Big|^2~.
\label{BRM12}
\end{equation}
So the Purity function is departing from unity and this is the consequence of creating entanglement between the two HOs.


\section{Quantum memory and collapsing - reviving}\label{Sec4}
To get more insight\textbf{,} let us investigate the entropy and Purity for a simple choice of GW: $\chi(t)=\chi_0\cos(\omega_g t - kX_3)$. For this choice, the HOs are oscillating along $X_1$ and $X_2$ directions respectively, while the GW is propagating along $X_3$. Here, $2\chi_0$ represents the amplitude of GWs and $\epsilon_+, \epsilon_{\times}$ represent two different types of polarizations of GWs. Furthermore, $\omega_g$ represents the angular frequency of GWs. Since we are interested in the dynamics (i.e. the time variation) of the entanglement generation, without any loss of generality, let us set $X_3=0$.  Then we have 
$\gamma(t)=-(1/2)\omega_{g}\chi_0\epsilon_+\sin(\omega_g t)$ and $\delta(t)=-\omega_{g}\chi_0\epsilon_{\times}\sin(\omega_g t)$,
and hence $\sigma(t)=\pm\hbar\chi_0\omega_g (\epsilon_+^2+\epsilon_{\times}^2)^{1/2} \sin(\omega_gt)$. Use of it yields
\begin{eqnarray}
&&c_+(t) = \pm \omega_g \chi_0\epsilon_0 \int_0^t dt_1 \sin(\omega_g t_1) e^{i\Omega t_1}
\nonumber
\\
&&= \pm i\omega_{g}\chi_0\epsilon_{0}\Big[\frac{ \sin(\frac{\Omega_+t}{2}) } {\Omega_+} e^{\frac{i\Omega_+t}{2}} -\frac{ \sin(\frac{\Omega_-t}{2})} {\Omega_-}e^{-\frac{i\Omega_-t}{2}}\Big]~~~~.
\end{eqnarray}
In the above\textbf{,} we defined $\epsilon_{0}={\sqrt{\epsilon_+^2+\epsilon_{\times}^2}}=1$, $\Omega_+=\omega_g+\Omega$ and $\Omega_-=\omega_g-\Omega$. With this\textbf{,} one finds
\begin{eqnarray}
&&\Big|c_+(t)\Big|^2=\Big(\frac{\chi_0\epsilon_{0}f}{1-f^{2}}\Big)^2 \Big[(1+f^{2})-(1-f^{2})\cos^{2}(\omega_{g}t)
\nonumber
\\
&&-2f^{2}\cos(\omega_{g}t )\cos(\Omega t)-2f \sin(\omega_{g}t) \sin(\Omega t)\Big]
\nonumber
\\
&&=\Big(\frac{\chi_0\epsilon_{0}f}{1-f^{2}}\Big)^2 \Big[(1+f^{2})-(1-f^{2})\cos^{2}(x)
\nonumber
\\
&&-2f^{2}\cos(x)\cos(\frac{x}{f})-2f \sin(x) \sin(\frac{x}{f})\Big]~,
\label{BRM13}
\end{eqnarray}
where $f=\frac{\omega_{g}}{\Omega}$ and $x=\omega_g t$.
We plot $y=\frac{\Big|c_+(t)\Big|^2}{\Big(\frac{\chi_0\epsilon_{0}f}{1-f^{2}}\Big)^2}$ Vs $x$ for a fixed value of $f$ ($f= 0.5$ and  $f=0.8$) and $z=\Big|\frac{\sigma(t)}{\hbar\chi_0\omega_g\epsilon_0}\Big|= \sin x$ Vs $x$, simultaneously.

Now we can study the following two cases.

{\bf Case I: Vanishing interaction} --
The interaction between the GWs and the HOs vanishes (i.e. $\sigma(t=T_m) = 0$) at different times, given by $\omega_gT_m = m\pi$, where $m$ is a positive integer (zero and negative integers are excluded as we are not considering at the initial time and time is in the forward direction). At these times\textbf{,} (\ref{BRM13}) yields
\begin{equation}
\Big|c_+(T_m)\Big|^2 = 2\Big(\frac{\chi_0\epsilon_{0}f^2}{1-f^{2}}\Big)^2\Big[ 1 - (-1)^m \cos\Big(\frac{m\pi}{f}\Big)\Big]~.
\label{BRM14}
\end{equation}
Note that, depending on the values of the parameters, the above one can be non-vanishing. This indicates that if the interaction is switched on between time $t=0$ to $t=T_m$, the modification to the initial quantum state retains even at the time ($t=T_m$) when the interaction with GWs is switched off. Therefore we will find that the initial non-entangled state, once entangled due to interaction, remains in the entangled state even at the point of time when interaction went off. This is quite clear from the figures. This phenomenon is ``similar'' to line of classical memory effect and therefore we call it as ``quantum memory'' effect.  
\begin{figure}[h!]
\includegraphics[scale=0.8]{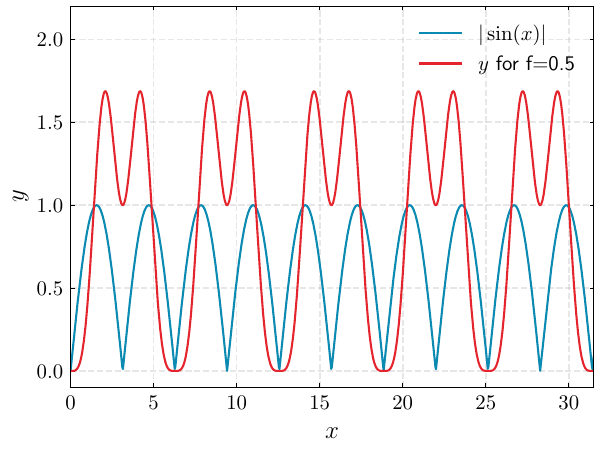}
\caption{Blue color curve represents $|\sin x|$ Vs $x$ plot while red one is for $y$ vs $x$ with the choice $f=0.5$.}
\label{Fig1}
\end{figure}
 
\begin{figure}[h!]
\includegraphics[scale=0.58]{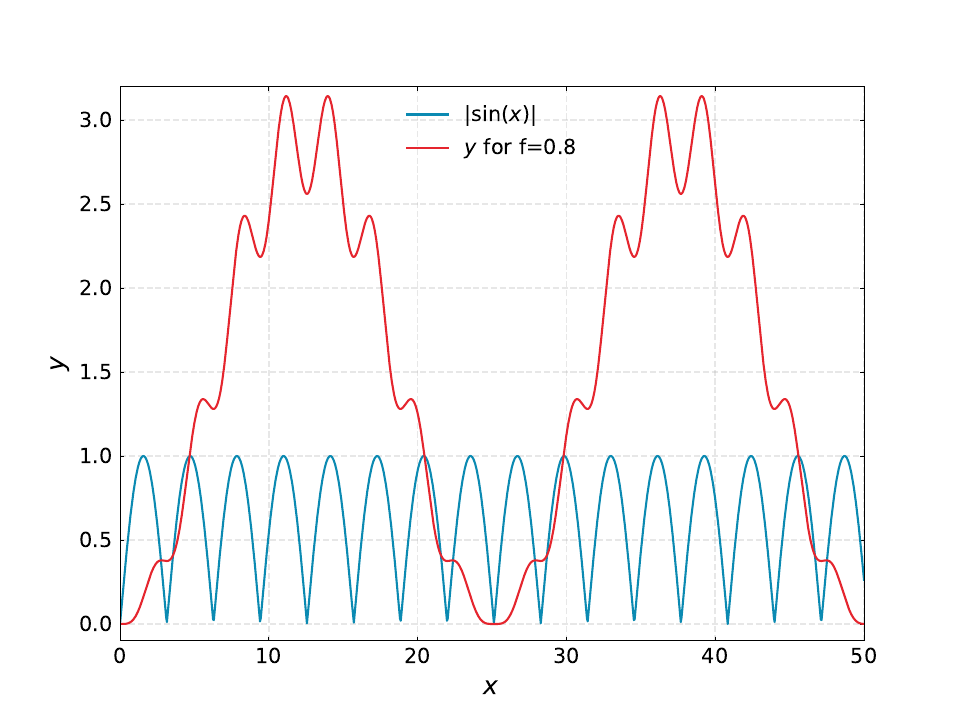}
\caption{Blue color curve is for $|\sin x|$ Vs $x$ while red one represents $y$ vs $x$ plot with the choice $f=0.8$.}
\label{Fig2}
\end{figure}

 {\bf Case II: Maximum interaction} --
 Another interesting feature is visible from the figures (particularly see Fig. \ref{Fig2}). For a bunch of maxima of the interaction, the value of $y$ is increasing and reaches at a maximum value. Then for the next few bunch of maxima, $y$ decreases as shown in the highlighted region of the FIG \ref{Fig2}. The same is happening periodically. This can be viewed as entanglement collapsing (decrease of $y$) and reviving (increase of $y$). 


\section{Effective thermality}\label{Sec6}
Further a very interesting analogy can be drawn between the density matrix (\ref{BRM9}) with the thermal density matrix of a harmonic oscillator.
If a one-dimensional harmonic oscillator with frequency $\Omega_0$ is at a temperature $\mathcal{T}$, then the density matrix is given by
\begin{equation}
\hat{\rho}_{th} = \Big(1-e^{-\frac{\hbar\Omega_0}{k_B\mathcal{T}}}\Big)\sum_{m=0}^\infty e^{-\frac{m\hbar\Omega_0}{k_B\mathcal{T}}} \ket{m,t=0}\bra{m,t=0}~.
\label{rhothm}
\end{equation}
In the above, $k_B$ is the Boltzmann constant. Here the frequency has been chosen $\Omega_0$, because our HOs have the same frequency (see Eq. (\ref{BP1})).
Then at a particular instant in time ($t$), a comparison between (\ref{BRM9}) and (\ref{rhothm}) reveals a remarkable observation, establishing a connection between the interaction of gravitational waves (GWs) and the temperature of the bath. The interaction with GWs seems to have an analogy with the thermal state of free HO with frequency $\Omega_0$, if one identifies
\begin{equation}
\Big|\frac{A_2}{A_1}\Big|^2=|c_+|^2=e^{-\frac{\hbar\Omega_0}{k_B\mathcal{T}}}~.
\label{GWTH}
\end{equation}
This relationship enables us to identify an effective temperature as
\begin{equation}
\mathcal{T}(t)=-{\frac{\hbar\Omega_0}{k_B\ln\Big|\frac{A_2}{A_1}\Big|^2}}=-{\frac{\hbar\Omega_0}{k_B\ln |c_+(t)|^2}}~.
\label{BP2}
\end{equation}
Additionally, the mean value of the occupation number of each HO, is given by
\begin{equation}
\mathcal{N}_1(\Omega_0,t)=\frac{1}{e^{\frac{\hbar\Omega_{0}}{k_{B}\mathcal{T}(t)}}-1}~.
\end{equation}
This observation strongly suggests that in the presence of gravitational interactions, our HOs not only get entangled but also any one of the HOs can effectively feel a thermal bath with a temperature, given by (\ref{BP2}). Consequently, it enhances the detection capability of GWs. When GWs traverse the LIGO detector, each individual arm of the LIGO detector which were oscillating, now register a finite positive effective temperature. This theoretical insight implies that GWs leave distinct signatures on the detector system as they pass through, potentially paving the way for experimental GW detection in the near future.

Note that $\mathcal{T}$ is time dependent and has been derived using an analogy with a HO in thermal bath which is at equilibrium. This poses a question on the validity of such analogy.  
Since our actual situation is time evolving one, in principle we need to invoke “non-equilibrium” definitions of thermodynamic quantities. So far the formalism is not well shaped. However there are few suggestions and advancements. In the non-equilibrium steady states, various thermometers, sensitive to different degrees of freedom (DOF), will show different temperatures. This leads to more than one temperature \cite{R1}. Consequently, the equilibrium version of the zeroth law is not applicable in global sense. In this case the validation of zeroth law can be restricted within those DOF, which defines one particular temperature. For example, suppose a system is composed of two types of DOF. Consequently one can define two values of temperature for these respective DOF. Then the zeroth law is valid within that particular DOF. This gives rise to different notions of  ``local'' temperature for the whole system. Here our motivation for identification is similar in nature. At each instant of time, the corresponding DOF gives rise to temperature (\ref{BP2}) which is ``local'' in time. However we then encounter heat flux, temperature gradient etc., among different DOF (see discussion in Section 4.1 of \cite{R1} for details). S. Weinberg suggested a way to define temperature to encounter the non-equilibrium situations. This is given by $\Big|\mathcal{N}_1(\Omega_0)/\mathcal{N}_1(-\Omega_0)\Big| = \exp[-(\hbar\Omega_0)/(k_B \mathcal{T})]$ (see discussion in Section 6.2 of \cite{R1} for details). Here we adopted the same spirit.
However, the concept of the ``global'' equilibrium condition prevails when considering sufficiently low-frequency modes of gravitational waves. These low-frequency modes naturally generate adiabatic variations \cite{Celeghini:1991yv} in the purity function. Furthermore, the effective temperature consistently remains positively definite, as $|c_+|^2$ consistently stays below unity, owing to the weak interaction amplitude of gravitational waves.


\section{Discussion}\label{Sec5}
In this study, we focus on a system comprising an isotropic bi-dimensional oscillator interacting with linearly polarized gravitational waves, specifically within the transverse traceless gauge, encompassing both plus and cross polarizations. To explore the entanglement phenomenon, we adopted a rotated coordinate system, effectively aligning the plus polarization with the cross polarization in this particular basis. This investigation led to a captivating discovery: the interaction of gravitational waves (GWs) with two independent isotropic oscillators, moving perpendicular to each other, induces entanglement within the quantum realm. Our exploration using second-order perturbation theory revealed this captivating insight. Investigating the dynamics of entanglement unveiled two remarkable aspects: firstly, a lingering entanglement persists even after the GWs interaction ceases—an intriguing phenomenon we've labeled the ``quantum memory effect''. Secondly, this entanglement exhibits dynamic fluctuations, rising and falling with the peaks of the GWs interaction. 
It may be noted that the reason behind this quantum memory lies in the time scale of the purity function, which surpasses the time scale of external periodic gravitational waves. Interestingly, the time scale of the purity function is not solely determined by the frequency of gravitational waves but also relies on the frequency of our quantum mechanical oscillator. Another noticeable feature of the interaction of with GWs is that the HOs can effectively feel a thermal bath in which they are immersed. The temperature of the bath depends on the oscillation frequency of the HO and the properties of the GWs. 


Recent studies have indicated that Newtonian gravity, when interacting with an optomechanical system, can induce gravity-induced quantum entanglement at the second order of gravitational interaction \cite{PhysRevD.106.126005, PhysRevD.108.106014}. This is considered a key aspect of the quantum nature of the gravitational field. In contrast, our research reveals that the dynamic behavior of entanglement can be naturally explained by the interaction of {\it classical} gravitational waves with a mechanical oscillator system. This discovery may unveil various aspects of (classical)gravitational waves, providing a different perspective from previous studies \cite{Pikovski:2013qwa}.

Moreover, our theoretical model, meticulously crafted to emulate LIGO's arms, holds promise as an alternative laboratory to explore this entanglement phenomenon using GWs—an ongoing pursuit ripe with potential. This innovative approach not only opens avenues for GW detection but also promises insights into unraveling the enigmatic classical characteristics inherent in these waves. 
Few comments are in order.
\begin{itemize} 
\item Our focus remained on elucidating the quantum behavior of the detector. It is noteworthy that the detection of GWs occurs at a length scale around $\sim 10^{-18} m$ (see \cite{Caves:1980rv,Thorne:1997ut}), which is just below the atomic scale. Therefore, from a practical standpoint, the quantum nature of the detector model cannot be disregarded. Consequently, our study primarily examines the impact of classical GWs on the quantum detector to understand the implications at the interface of classical GWs and the quantum model of the detector. In addition, the wave length of GWs is larger than the detector's size so that the detector can be considered as a point particle. In this case the GWs are more classical than the detector. Therefore from this point of view consideration of detector as quantum while GWs as classical may be a reasonable approximation. However to have better illumination of quantumness of gravity, treating GWs as quantum operator will be more interesting. At this point we refrain from investigating such thing as the criterion of {\it system locality} (see discussion in \cite{PhysRevD.108.L101702}) is being violated for our interaction term and therefore exploration of quantum nature of gravity through entanglement will not be so reliable.
Exploring additional novel consequences arising from this interaction would be intriguing. For instance, treating the GWs themselves as quantized could lead to entanglement between the two modes of each oscillator arm. This phenomenon contrasts with classical scenarios where there is no classical information sharing between the two modes. Recently we explored this aspect in \cite{Nandi:2024jyf}.
\item Additionally, the coupling to fluctuating quantum fields, including scalar and electromagnetic fields, did not arise here. This exclusion is justified by the nature of the interaction between classical GWs and the quantum detector, which can be viewed as a classical fluctuation of a quantum detector. 
In-fact, we consider our oscillator detector as a neutral entity, meaning it doesn't directly interact with the background electromagnetic field. This approach allows us to disregard electromagnetic vacuum fluctuations and radiation pressure effects in our study, especially considering that low velocity oscillating detectors are associated with low-temperature effects. Note that the radiation pressure is proportional to fourth power of temperature. However, we do acknowledge the existence of the Higgs field as the sole real scalar field in nature. In our analysis the detector has been taken as quantum point particle with respect to GWs as the wavelength of GWs is very much larger than its size. However such approximation is not feasible with respect to excitation of  Higgs field as they are sub-atomic particles. Therefore the detector can not be taken as atomic size particle with respect to Higgs particle and hence consideration of quantum nature of detector with respect to Higgs will be questionable.
Therefore macroscopic objects with respect to Higgs, such as our oscillator's mass, can potentially interact with the Higgs field through their constituent particles. Then the interaction is sensitive to the substructure  of this oscillating objects. In this situation it is expected that the deviation of detector by Higgs will be negligible compared to that by GWs.
This is primarily due to the significantly larger masses of macroscopic oscillator objects compared to elementary particles that directly interact with the Higgs field.
So, our analysis specifically focuses on the deviation in arm caused by GWs. This focused approach is in perfect alignment with the primary goal of LIGO's detection capability. 
\item Finally, we believe that the nontrivial dynamics of entanglement induced by classical gravitational waves have not yet been extensively explored in the existing literature, to the best of our knowledge. We aim to delve deeper into these aspects in our future endeavors.
\end{itemize}

\vskip 5mm
\section{Acknowledgements}
One of the authors (PN) acknowledges the support from a 
postdoctoral fellowship grants from Stellenbosch University, South Africa. We also thank Sibasish Ghosh,  IMSc (Chennai, India) for giving various insights on entanglement. The anonymous referee is also greatly acknowledged for giving various important remarks.

\bibliographystyle{apsrev}
\bibliography{bibtexfile}


\end{document}